\documentclass[preprint,aps,prd,amsmath,amssymb]{revtex4}
\usepackage{graphicx}
\usepackage{color}
\newcommand{\beq}{\begin{eqnarray}}
\newcommand{\eeq}{\end{eqnarray}}

\newcommand{\bmp}{\noindent\begin{minipage}{16cm}}
\newcommand{\emp}{\end{minipage}\vskip 7mm} 

\usepackage{graphicx}
\usepackage{dcolumn}
\usepackage{bm}
\usepackage{amsmath}
\usepackage{amsfonts}
\usepackage{bbm}
\usepackage{subfigure}

\usepackage{ulem}

\def\drawbox#1#2{\hrule height#2pt
        \hbox{\vrule width#2pt height#1pt \kern#1pt
              \vrule width#2pt}
              \hrule height#2pt}

\def\Asym#1#2{\vcenter{\vbox{\drawbox{#1}{#2}
              \kern-#2pt 
              \drawbox{#1}{#2}}}}

\begin{document}
\title{{\large Gauge Coupling Unification \\ via \\ A  Novel Technicolor Model}}
\author{Sven Bjarke {\sc Gudnason}}
\email{gudnason@nbi.dk}
\author{Thomas A. {\sc Ryttov}}
\email{ryttov@nbi.dk}
\affiliation{The Niels Bohr Institute, DK-2100 Copenhagen \O, Denmark }
\author{Francesco {\sc Sannino}}
\email{sannino@nbi.dk}

\affiliation{University of Southern Denmark, DK-5230 Odense,}
\affiliation{The Niels Bohr Institute, DK-2100 Copenhagen \O, Denmark }

\begin{abstract}
We show that the recently proposed minimal walking technicolor theory together with a small modification of the Standard
Model fermionic matter content leads
to an excellent degree of unification of the gauge couplings. We compare the degree of unification with various time-honored technicolor models and the minimal supersymmetric extension of the Standard Model. We find that, at the one-loop level, the new theory provides a degree of unification higher than any of the other extensions
above.  The phenomenology of the present model is very rich with various potential dark matter candidates.
\end{abstract}


\maketitle
\section{Introduction}
Coupling unification of all forces is a milestone for any extension of the standard model. In this letter we
explore the unification issue for four dimensional extensions of the standard model (SM) in which the Higgs
sector is replaced by a technicolor-like mechanism \cite{TC}.

Much progress has been made recently in developing new models of technicolor type able to address the
old problems for technicolor
\cite{Sannino:2004qp,Hong:2004td,Evans:2005pu,Dietrich:2005jn,Dietrich:2005wk}.
We have even identified and carefully studied within these models
possible dark matter particles in the form of technibaryons
\cite{Gudnason:2006yj}. Our technicolor differs from the more
traditional models of technicolor used in the past since the
technifermions transform according to higher dimensional
representations of the underlying technicolor gauge group.  This does
not exclude the possibility that such fermions can be interpreted at
some higher energy as transforming according to the fundamental
representation of a higher rank
group. One of the key features
of these technicolor theories is that they {\it walk}
\cite{Holdom:1981rm,{Yamawaki:1985zg},{Appelquist:an},
  MY,Lane:1989ej,Cohen:1988sq}
for a very low number of technifermions and technicolors.
{Some of the
problems of the simplest technicolor models, such as a large contribution to the oblique parameters \cite{Peskin:1990zt}, are alleviated \cite{Appelquist:1998xf,Sundrum:1991rf,Appelquist:1999dq}
when considering new gauge dynamics in which the coupling does not run with respect to the energy scale but rather walks, i.e.~evolves very slowly \cite{Holdom:1981rm,{Yamawaki:1985zg},{Appelquist:an}, MY,Lane:1989ej}.} The simplest of
such models which passes the electroweak precision
tests, requires the technicolor matter to transform according to
the two index representation \footnote{The adjoint representation of $SU(2)$ - technicolor.}
of the technicolor gauge group
\cite{Sannino:2004qp,Hong:2004td,Dietrich:2005jn,Dietrich:2005wk}. In
\cite{Dietrich:2006cm}, the reader
will find an exhaustive analysis of asymptotically free
non-supersymmetric gauge theories
with fermions in a given arbitrary
higher dimensional representation of the $SU(N)$ gauge group and their
use to dynamically break the electroweak symmetry.

The minimal walking technicolor (MWT) model has
been introduced in \cite{Dietrich:2005jn,Dietrich:2005wk,Gudnason:2006yj}  and consists
of a two technicolor gauge theory with technifermions in the
two-index symmetric (i.e.~adjoint)
representation of the technicolor gauge group.  To avoid Witten's
global $SU(2)$ anomaly \cite{Witten:fp}, one introduces a new lepton
family.

 We start by investigating the  one-loop evolution of the SM couplings
 once the SM Higgs is replaced by the MWT model. Quite surprisingly we
 find that the SM coupling constants unify much better than with the
 standard model Higgs being
 present. We compare our results with
 different time-honored technicolor models and show that either they
 are not competitive unification wise or they are not
 a prime candidate
 for walking technicolor theories according to recent results
 \cite{Dietrich:2006cm}. With a small modification of the technicolor
 gauge interactions we show how we can envision unification also
 with the technicolor coupling constant at the same energy scale.

Technicolor requires some other mechanism to provide the standard model fermion masses. This mechanism could have an effect on our results. We have estimated these corrections by providing a simple/minimal model which consists in adding a new Higgs field on the top of the minimal walking theory whose main purpose is to provide mass to standard model fermions. This construction has already been used in the literature
\cite{Simmons:1988fu,Dine:1990jd,Kagan:1990az,Kagan:1991gh,Carone:1992rh,Carone:1993xc}.
 In a more natural theory this field will be replaced, perhaps, by some new strong dynamics. The model parametrizes our ignorance of a more fundamental extended technicolor theory.  Surprisingly we find a degree of unification higher than in the theory without a mechanism for fermion mass generation. This result is very encouraging.

A general feature of a unified theory of the SM interactions is the prediction of the proton decay. A unification energy scale of the order of, or larger than, $10^{15}$ GeV leads, typically, to phenomenologically acceptable proton decay rates.
Despite the good, but yet not
  perfect, degree of
  unification - when compared, for example, to the minimal
    supersymmetric standard model result for unification - we discover
  that the proton decays too fast since the unification scale is
quite low.
To cure the proton decay problem we then add a QCD colored Weyl fermion
  transforming according to the adjoint representation of $SU(3)$ and
  one Weyl fermion transforming according to the adjoint of
  $SU_L(2)$. These fermions are known in supersymmetric extensions of
  the standard model as the gluino and the wino.  We then compare with
  the supersymmetric predictions for unification at the same order in
  perturbation theory and discover that the present theory unifies
  better. Since we are not insisting on
  supersymmetry, there is no
  reason to expect the introduced fermions to be degenerate with the
  associated gauge bosons. Unification and naturality also suggest the
  presence of a Weyl fermion associated to the hypercharge gauge
  interaction and hence this degree of freedom is added in the model
  (the bino).

The reader may consider adding matter transforming according to even
higher dimensional representations than the adjoint one or more
generally higher than the two-index type matter. We know
\cite{Dietrich:2006cm}, however, that a very limited number of
theories with fermions in higher dimensional representations remain
asymptotically free when the rank of the gauge group increases.  The
unified gauge group must necessarily have quite a large rank
constraining matter to have at most two indices to insure asymptotic
freedom. Recall, that to avoid low energy fine tuning of the
coupling constants the unified gauge theory must be asymptotically
free. This requirement, {\it de facto}, limits the maximum allowed
representation in the theory.

The phenomenology, both for collider experiments and cosmology, of this
novel extension of the standard model is very rich with many features
common to both supersymmetry and technicolor.


\section{Notation and Conventions}
The evolution of the coupling constant $\alpha_n$, at the
one-loop level, of a gauge theory
is controlled by
\begin{equation}\label{running}
{\alpha_{n}^{-1}(\mu) = \alpha_{n}^{-1}(M_Z) - \frac{b_n}{2\pi}\ln
\left(\frac{\mu}{M_Z}\right) \ ,}
\end{equation}
where $n$ refers to the gauge group being $SU(n),$ for $n\geq 2$ or $U(1),$
 for $n=1\ $.

The first coefficient of the beta function $b_n$ is
\begin{eqnarray}
b_n =  \frac{2}{3}T(R) N_{wf} + \frac{1}{3} T(R')N_{cb} -
\frac{11}{3}C_2(G) \ ,
\end{eqnarray}
where $T(R)$ is the Casimir\footnote{Note that here we are using a different normalization than the one adopted in \cite{Dietrich:2006cm}.} of the
representation $R$ to which the
fermions belong, $T(R')$ is the
Casimir of the representation $R'$
to which the bosons belong. $N_{wf}$ and $N_{cb}$ are respectively the
number of Weyl fermions and the number of complex scalar bosons. $C_2(G)$ is the
quadratic Casimir of the adjoint representation of the gauge group.

The SM gauge group is $SU(3)\times SU(2)\times U(1)$. We have three
associated coupling constants which one can imagine to unify at some
very high energy scale $M_{GUT}$. This means that the three
couplings are all equal at the scale $M_{GUT}$, i.e.
$\alpha_3(M_{GUT})= \alpha_2(M_{GUT})=\alpha_1(M_{GUT})$ with
$\alpha_1= \alpha/(c^2\cos^2 \theta_w)$ and $\alpha_2 = \alpha/
\sin^2\theta_w$, where $c$ is a normalization constant to be
determined shortly.

Assuming one-loop unification using
Eq.~(\ref{running}) for $n=1,2,3,$ one
finds the following relation
\begin{eqnarray} \label{unification}
\frac{b_3-b_2}{b_2-b_1} & = & \frac{\alpha_3^{-1} -\alpha^{-1}\sin^2
\theta_w}{(1+c^2)\alpha^{-1} \label{1}
\sin^2\theta_w-c^2\alpha^{-1}} \ .
\end{eqnarray}
In the above expressions the
Weinberg angle
$\theta_w$, the electromagnetic coupling constant $\alpha$ and the
strong coupling constant $\alpha_3$ are all evaluated at the
$Z$ mass. For a
given particle content we shall denote the LHS of
Eq.~(\ref{unification}) by $B_{\rm theory}$ and the RHS by
$B_{\rm
  exp}$. Whether $B_{\rm theory}$ and $B_{\rm exp}$ agree is a simple way to
check if
the coupling constants unify. We shall use the experimental
values $\sin^2 \theta_w (M_Z) = 0.23150\pm 0.00016$,
$\alpha^{-1}(M_Z) = 128.936 \pm 0.0049$, $\alpha_3(M_Z) =
0.119\pm 0.003$ and $M_Z = 91.1876(21)$ GeV \cite{Yao:2006px}.
The unification scale is given by the expression
\begin{eqnarray}
{M_{GUT} = M_Z
\exp
\left[{{2\pi}\frac{\alpha_2^{-1}(M_Z)-\alpha_1^{-1}(M_Z)}{b_2-b_1}}\right]
\ . }
\end{eqnarray}

While the normalizations of the
coupling constants of the two
non-Abelian gauge groups are fixed by
the appropriately normalized
generators of the gauge groups, the
normalization of the Abelian
coupling constant is a priori arbitrary. The normalization of the
Abelian coupling constant can be
fixed by a rescaling of the
hypercharge $Y \rightarrow cY$ along with
$g\to g/c\ $. The normalization
constant $c$ is
chosen by imposing that all three coupling constants have a common
normalization
\begin{eqnarray}
\text{Tr}\,(c^2Y^2) = \text{Tr}\,(T_3^2) \ ,
\end{eqnarray}
where $T_3$ is the generator of the weak
  isospin and
the trace is over all the relevant fermionic particles on which the
generators act. It is sufficient to fix it for a given fermion
  generation (in a complete multiplet of the unification group).

The previous normalization is consistent with
an $SU(5)$-type normalization for the generators of $U(1)$
of hypercharge,
$SU(2)_L$ and $SU(3)_c\ $.

As well explained in the paper by Li
and Wu \cite{Li:2003zh}:
{\it  At one-loop a contribution to $b_3 - b_2$ or $b_2 - b_1$ emerges
  only from particles not forming complete representations
  \footnote{Such as the five and the ten dimensional representation of
    the unifying gauge group $SU(5)$.} of the unified gauge
  group}. {}For example the gluons, the weak gauge bosons and the
Higgs particle of the SM do not form complete
representations of $SU(5)$ but ordinary quarks and leptons do. Here we
mean that these particles form complete representations of $SU(5)$, all
the way from the unification scale
    down to the electroweak scale.
The particles not forming complete representations will presumably
join at the unification scale with new particles and together then
form complete representations of the unified gauge group. Note, that
although there is no contribution to the unification point of the
particles forming complete
representations, the running of each
coupling constant is affected by all of the particles present at low
energy.

As a warm up, we consider the SM with
$N_g$ generations. In
this case we find $c=\sqrt{3/5}$, which is the same value one finds
when the hypercharge is upgraded to one of the generators of $SU(5)$,
and therefore the beta function
coefficients are

\begin{eqnarray}
b_3 & = & \frac{4}{3}N_g -11 \ ,\\
b_2 & = & {\frac{4}{3} N_g - \frac{22}{3} +
\underbrace{\frac{1}{6}}_{\rm Higgs} \ , } \\
b_1 & = & {\frac{3}{5}\left( \frac{20}{9}N_g +\frac{1}{6} \right) =
\frac{4}{3}N_g +\underbrace{\frac{1}{10}}_{\rm Higgs} \ .}
\end{eqnarray}

Here $N_g$ is the number of
generations. It is clear that the
SM does not unify since $B_{\rm theory} \sim 0.53$ while
$B_{\rm exp} \sim 0.72\ $.

\begin{figure}[htbp]
\begin{center}
\includegraphics[width=0.6\linewidth]{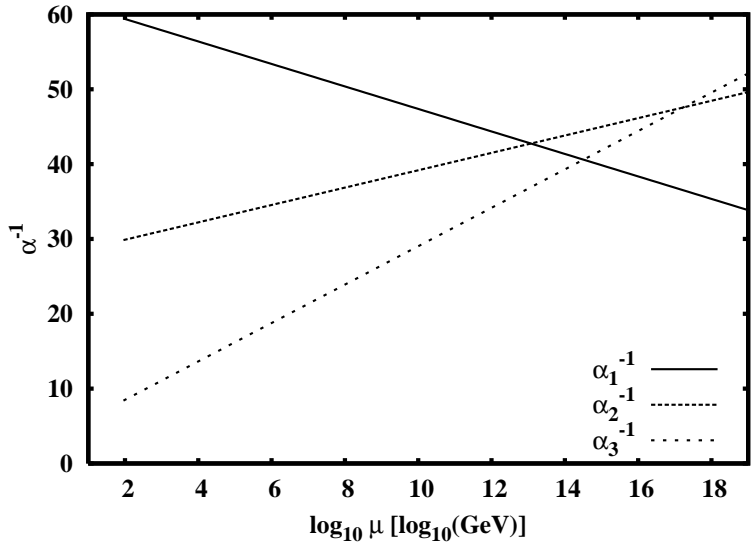}
\end{center}
\caption{The running of the three standard model gauge couplings.}
\label{SM}
\end{figure}

Note that the spectrum relevant for computing
$B_{\rm theory}$  is constituted by the gauge bosons and the standard
model
Higgs. The contribution of quarks and leptons drops out in agreement
with the fact that they form complete representations of the unifying
gauge group which, given the present normalization for $c$, is at
least $SU(5)$. Hence the predicted value of $B_{\rm theory}$ is
independent of the number of
generations. However the overall running
for the three couplings is dependent on the number of
  generations and in
Fig.~\ref{SM} we show the behavior of the three couplings with $N_g
= 3$.

\section{Studying SU(3)$\times$SU(2)$\times$U(1) Unification in Technicolor}
Here we compare a few examples in which the standard model Higgs is
replaced by a technicolor-like theory. A similar analysis was performed in \cite{Christensen:2005bt}. In this section we press on recent
phenomenological successful technicolor models with technimatter in
higher dimensional representations and demonstrate that the simplest
model helps unifying the SM couplings while other more
traditional approaches are less successful. We also show that by a
small modification of the technicolor
dynamics, all of the four
couplings can unify \footnote{Since the technicolor dynamics is
  strongly coupled at the electroweak scale the last point on the
  unification of all of the couplings is meant to be only
  illustrative.}.

\subsection{Minimal Walking Technicolor (MWT)}
\begin{figure}[!tbp]
\begin{center}
\includegraphics[scale=0.6]{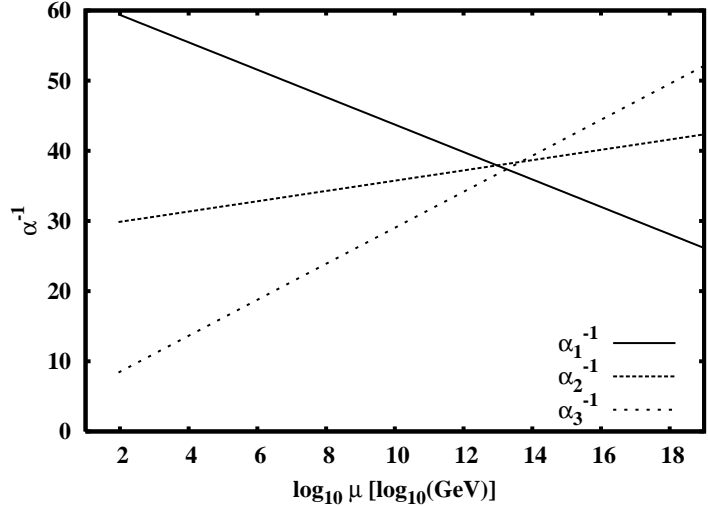}
\end{center}
\caption{The running of the SM gauge couplings in the
    presence of adjoint technifermions (the technicolor
coupling is not included here).}\label{TC}
\end{figure}

We examine what happens to the running of the SM couplings when the Higgs sector is replaced by the
MWT theory proposed and investigated in  \cite{Dietrich:2005jn,Dietrich:2005wk}.
This model has
technicolor group $SU(2)$ with two techniflavors in the two-index
symmetric representation of the technicolor group. As already
mentioned to avoid Witten's $SU(2)$
anomaly, the minimal solution is to
add a
new lepton family. Gauge anomaly cancellation as well as the
previously mentioned anomaly do not uniquely fix the hypercharge
assignment of the theory. Here we start by taking the simplest one in
which the new techniparticles have electric $Q$ and hypercharge $Y$ mimicking the ones of the ordinary quarks and leptons
\begin{align}
T_{L}^{(Q, Y)} &= \binom{U_L^{\frac{2}{3},
\frac{1}{6}}}{D_L^{-\frac{1}{3},\frac{1}{6}}} \ , \qquad
&T_{R}^{(Q,Y)} &= \left( U_{R}^{\frac{2}{3},\frac{2}{3}},
D_{R}^{-\frac{1}{3},-\frac{1}{3}} \right) \ ,\\
 \mathcal{L}_{L}^{(Q,Y)} &= \binom{\nu_{L,\zeta}^{0,-\frac{1}{2}}}{
   \zeta_L^{-1,
 -\frac{1}{2}}} \ , \qquad &\mathcal{L}_R^{(Q,Y)} &= \left(
 \nu_{R,\zeta}^{0,0},
 \zeta_{R}^{-1,-1}  \right) \ .
\end{align}
We still assume an $SU(5)$-type
unification leading to $c^2=3/5$.
The beta function
coefficients will be those of
  the SM minus the Higgs plus the extra
contributions from the techniparticles, ergo
\begin{eqnarray}
b_3 & = & \frac{4}{3}N_g -11 \ ,\\
b_2 & = & \frac{4}{3} N_g - \frac{22}{3} + \frac{2}{3}
\frac{1}{2}\left( \frac{2(2+1)}{2} + 1 \right) = \frac{4}{3} \left(
N_g + 1 \right) - \frac{22}{3} \ ,\\
b_1 & = & \frac{3}{5} \left( \frac{20}{9} N_g + \frac{20}{9} \right)
= \frac{4}{3}\left( N_g+1 \right) \ ,
\end{eqnarray}
where $N_g$ is the number of ordinary SM generations. From this we
see that $B_{\rm theory}=0.68$ and $B_{\rm exp} = 0.72\ $ and hence argue that we have a better
unification than in the standard model
with an elementary Higgs. The running of the SM couplings is shown in
Fig.~\ref{TC} for three ordinary standard model generations.
Note that the increase in $B_{\rm theory}$ with respect to the
SM is due to the fact that,
typically, bosonic contributions
are numerically suppressed with respect to fermionic
ones and
that,
while $b_1-b_2 = 22/3$ receives only a contribution from the gauge
sector, $b_2-b_3 = 11/3 + 4/3$ has
two contributions, a gauge one and a
fermionic one. These results are a
direct consequence of the fact that we have no
ordinary quarks related to the new leptonic family.

\subsection{Traditional Walking and Non-Walking One Family Model}
Here the technicolor particles also carry ordinary color and the
technifermions constitute complete representations of $SU(5)$, hence
the SM coupling unification receives no improvement with respect to
the SM case. This is so since the numerical effect of the Higgs on
the unification is small. {}For a one-family $SU(N)$ theory, we
  have $B_{\rm theory} = 1/2 \ $.

\subsection{Partially Electroweak-Gauged Technicolor}

This approach consists in letting
only one doublet of techniquarks transform non-trivially under the electroweak
symmetries with the rest of the matter remaining in electroweak singlets, as suggested in
\cite{Dietrich:2005jn} and later on used in
\cite{Christensen:2005cb}. In this case all techniquarks transform
still under the technicolor gauge group and hence contribute to
rendering
the technicolor dynamics quasiconformal without affecting the leading
perturbative contribution  to the electroweak precision
parameters. The reader can find a rather exhaustive investigation of
walking technicolor theories with fermions in different
representations of the technicolor gauge group in
\cite{Dietrich:2006cm}.

Denoting by ${\rm
  dim}[R_T]$ the dimension of the technicolor representation under
which all the techniquarks transform we find the following result for
$B_{\rm theory}$
\begin{eqnarray}
B_{\rm theory} = \frac{1}{2}\left[\frac{11+{\rm dim}[R_T]}{11-{\rm dim}[R_T]/ {5}}\right] \ .
\end{eqnarray}
Since we have gauged only one technidoublet with respect to the weak interactions to avoid Witten's global anomaly with respect to the weak interactions we take ${\rm dim}[R_T]$ to be even. We have used the following hypercharge assignment free from gauge anomalies
\begin{align}
Y(T^a_L)=& \; 0 \ ,&\quad Y(U^a_R,D^a_R)&=\left(\frac{1}{2},-\frac{1}{2}\right) \ ,&\quad a=1,\dots,{\rm dim}[R_T]
\ ,
\end{align}
for the technidoublet charged under
the electroweak interactions.

We find that $B_{\rm theory}$ is 0.73 for ${\rm dim}[R_T]=4$ and it
increases with larger values of ${\rm dim}[R_T]\ $. We can also consider
the case ${\rm dim}[R_T]$ odd while solving Witten's anomaly by adding
for example a new weak doublet uncharged under technicolor. We see
that from the unification point of view partially electroweak gauged
technicolor models are comparable with the MWT model presented
earlier.

However, for the model to be phenomenologically viable the new
technicolor theory should pass the electroweak precision
constraints. A complete list of
walking-type technicolor theories
passing the precision tests can be found in
\cite{Dietrich:2006cm}. The simplest unification condition requires
the technicolor representation, in this case, to be
four dimensional. This can only be achieved when the technifermions
are arranged in the fundamental representation of the
$SU(4)$-technicolor gauge group. According to Table III in
\cite{Dietrich:2006cm} one needs, at least, fifteen techniflavors for
the theory to have a walking behavior with a reasonable
$S$
parameter. In \cite{Dietrich:2006cm},
this theory has not been listed
as a prime candidate and hence will
not be considered further here.

\subsection{The Technicolor Coupling Constant}

Until now we have not discussed the technicolor coupling
constant $\alpha_{TC}$.
It is possible that the technicolor
interaction does not unify with the other three forces or unifies
later. A single step unification is though esthetically more appealing to
us. Here we focus on the minimal walking theory which has already
shown to be a promising theory for the unification of the standard
model couplings.
Remembering that the Casimir of the
two-index symmetric representation of $SU(N_{TC})$ is
$(N_{TC}+2)/2$ the first coefficient
of the beta function
$b_{TC}$ is easily found to be
\begin{eqnarray}
b_{TC} = \frac{2}{3} (N_{TC}+2)  N_{f} - \frac{11}{3}N_{TC} \ ,
\end{eqnarray}
where $N_{TC}$ is the number of
technicolors and $N_f$ is the
number of techniflavors. For two
colors and two flavors we find
$b_{TC}=-2$. Observing that, somewhat accidentally, also $b_2=-2$ for
three ordinary SM
generations, we
conclude that the technicolor coupling constant cannot unify with the
other three couplings at the same point. We are assuming, quite
naturally, that the low energy starting points of $\alpha_2$ and
$\alpha_{TC}$ are different.

Insisting that the technicolor coupling constant must
unify with the other coupling constants at
$M_{GUT}$, we need to modify
at a given scale
$ {X}<M_{GUT}$ either the
overall running of the SM couplings or the one of technicolor. To make less
steep the running of the SM couplings one could add new
generations. To avoid the loss of asymptotic freedom for the week
coupling we find that at most only one entire new SM like
generation can be added at an intermediate
scale.
If we, however, choose not to modify the running of the SM coupling constants, the
running of the technicolor coupling constant must at some point $X < M_{GUT}$ become
steeper. This can be achieved by enhancing the number of technigluons
and lowering the contribution due to the
techniquarks at the
scale $X$. An elegant way to implement this idea is to imagine
that the techniquarks - belonging to
the three dimensional
two-index symmetric representation of $SU(2)$ - are
embedded in the
fundamental representation of $SU(3)$ at the scale $X$. At energies
below $X$ we have $b^{<
  X}_{TC}=-2$ and for energies larger than
$X$ we have $b_{TC}^{> X} = -29/3$. If we take the technicolor
coupling to start running at the electroweak scale $M_{EW}\sim 246 \
\text{GeV}$ and unifying with the three SM couplings at the
unification scale we find an expression for the intermediate scale
$X$
\begin{equation}
\ln X = \frac{1}{b_{TC}^{< X} - b_{TC}^{>
 X}}\bigg\{2\pi \bigg(\alpha_{TC}^{-1}(M_{EW}) - \alpha_{TC}^{-1}(M_{GUT})\bigg) +
 b_{TC}^{< X} \ln M_{EW} - b_{TC}^{> X}\ln M_{GUT} \bigg\} \ .
 \label{XX}
\end{equation}
If we take the starting point of the running of the technicolor
coupling to be the critical coupling close to the conformal
window
we have ${\alpha_{TC}(M_{EW}) =
\pi/(3C_2(\square\hspace{-1pt}\square)) = \pi/6}\ $. Also
using the numbers
$\alpha_{TC}(M_{GUT}) = {\alpha_i(M_{GUT}) \sim 0.026\ ,
  i=1,2,3\ }$, $M_{GUT} \sim 9.45 \times 10^{12}
\,\text{GeV}$ we find the intermediate scale
to be $X \sim 830$ GeV.

\subsection{Proton Decay}
Grand Unified Theories lead, generally, to proton decay. Gauge bosons
of mass ${M_{V} < M_{GUT}}$ are
responsible for the decay of the proton into $\pi^0$ and $e^+$. The
lifetime of the proton is estimated to be \cite{Giudice:2004tc}
\begin{eqnarray}
\tau &=& {\frac{4f^2_{\pi}M^4_{V}}{\pi m_p \alpha_{GUT}^2 \left(1+D+F\right)^2
\alpha_N^2\left[A_R^2+\left(1+|V_{ud}|^2 \right)^2A_L^2\right]}} \\
& = &{\left( \frac{M_{GUT}}{10^{16}\ \text{GeV}} \right)^4 \left(
\frac{\alpha_{GUT}^{-1}}{35} \right)^2 \left( \frac{0.015\
\text{GeV}^3}{\alpha_N} \right)^2 \left(\frac{2}{\mathcal{A}} \right)^2 2.7
\times 10^{35}\ \text{yr} \ , }
\end{eqnarray}
where we have used $f_{\pi} = 0.131\ \text{GeV}$, the chiral
Lagrangian factor $1+D+F = 2.25$, the operator renormalization
factors $\mathcal{A} \equiv A_L=A_R$ and the hadronic matrix element
is taken from lattice results \cite{Aoki:1999tw} to be $\alpha_N
=-0.015\ \text{GeV}^3$. Following Ross \cite{Ross:1985ai}, we have
estimated $\mathcal{A} \sim 2$ but a larger value $\sim 5$ is quoted in \cite{Giudice:2004tc} .
The lower bound on the unification scale comes from
the Super-Kamiokande limit $\tau > 5.3 \times 10^{33}$
{yr}
\cite{Suzuki:2001rb}
\begin{eqnarray}
M_{GUT} > M_{V} & > & {\left( \frac{35}{\alpha_{GUT}^{-1}} \right)^{1/2}
\left( \frac{\alpha_N}{-0.015\ \text{GeV}^3}\right)^{1/2} \left(
\frac{\mathcal{A}}{2} \right)^{1/2}\ 3.7\times 10^{15}\ \text{GeV} \ .}
\end{eqnarray}

In the MWT model extension of the SM we find ${\alpha^{-1}_{GUT} \sim
  37.5}$ and $M_{GUT} \sim  10^{13} \,\text{GeV}$ yielding too fast
proton decay.

\subsection{Constructing a Simple Unifying Group}
We provide a simple embedding of our matter content into a unifying gauge group. To construct this group we first summarize the charge assignments in table \ref{single}.
\begin{table}[h]
\caption{Quantum Numbers of the MWT + One SM Family}
\begin{center}
\begin{tabular}{c|c|c|c|c}
&$SO_{TC}(3)$&$SU_c(3)$&$SU_L(2)$&$U_Y(1)$ \\
\hline \hline
$q_L$ &1&3&2&1/6 \\
$u_R$ & 1 &3 &1 & 2/3 \\
$d_R$ & 1&3 &1&-1/3\\
$L $& 1&1&2&-1/2 \\
$e_R$ & 1 &1&1&-1 \\
\hline
$T_L$ &3&1&2&1/6 \\
$U_R$ & 3 &1 &1 & 2/3 \\
$D_R$ & 3&1 &1&-1/3\\
${\cal L}_L $& 1&1&2&-1/2 \\
$\zeta_R$ & 1 &1&1&-1 \\
\end{tabular}
\end{center}
\label{single}
\end{table}
For simplicity we have considered right transforming leptons only
for the charged ones. Also, the techniquarks are classified as being
fundamentals of $SO(3)$ rather than adjoint of $SU(2)$. Except for
topological differences, linked to the center group of the two
groups, there is no other difference. This choice allows us to show
the resemblance of the technicolor fermions with ordinary quarks. We
can now immediately arrange each SM family within an ordinary
$SU(5)$ gauge theory. The relevant question is how to incorporate
the technicolor sector (here we mean also the new Lepton family). An
easy way out is to double the weak and hypercharge gauge groups as
described in table \ref{tabledouble}.
\begin{table}[h]
\caption{MWT + One SM Family enlarged gauge group}
\begin{center}
\begin{tabular}{c|c|c|c|c|c|c}
&$SO_{TC}(3)$&$SU_1(2)$&$U_1(1)$&$SU_c(3)$&$SU_2(2)$&$U_{2}(1)$ \\
\hline \hline
$q_L$ &1&1&0&3&2&1/6 \\
$u_R$& 1&1&0 &3 &1 & 2/3 \\
$d_R$ &1&1& 0&3 &1&-1/3\\
$L $&1&1 &0&1&2&-1/2 \\
$e_R$ &1&1& 0 &1&1&-1 \\
\hline
$T_L$ &3&2&1/6&1&1&0 \\
$U_R$ & 3 &1 & 2/3&1&1&0 \\
$D_R$ & 3&1 &-1/3&1&1&0\\
${\cal L}_L $& 1&2&-1/2&1&1&0 \\
$\zeta_R$ & 1 &1&-1&1&1&0 \\
\end{tabular}
\end{center}
\label{tabledouble}
\end{table}
This assignment allows us to arrange the low energy matter  fields
into complete representations of $SU(5)\times SU(5)$. To recover the
low energy assignment one invokes a spontaneous breaking of the
group down to $SO(3)_{TC}\times SU_c(3)\times SU_L(2) \times U_Y(1)$
\footnote{To achieve such as a spontaneous breaking of the gauge
group one needs new matter fields around or slightly above the grand
unified scale transforming with respect to both the gauge groups. }.
We summarize in table \ref{GUT} the technicolor and SM fermions
transformation properties with respect to the grand unified group.
\begin{table}[h]
\caption{GUT}
\begin{center}
\begin{tabular}{c|c|c}
&$SU(5)$&$SU(5)$ \\
\hline \hline
$\bar{A}_{SM}$ &1&$\overline{10}$\\
$F_{SM}$ & 1&5 \\
\hline
$\bar{A}_{MWT}$ &$\overline{10}$&1\\
$F_{MWT}$& 5&1 \\
\end{tabular}
\end{center}
\label{GUT}
\end{table}
Here the fields $A$ and $F$ are standard Weyl fermions and the gauge
couplings of the two $SU(5)$ groups need to be the same. We have
shown here that it is easy to accommodate all of the matter fields
in a single semi-simple gauge group. This is a minimal embedding and
others can be envisioned. New fields must be present at the grand
unified scale (and hence will not affect the running at low energy)
guaranteeing the desired symmetry breaking pattern.

\subsection{Providing Mass to the fermions}
We have not yet considered the problem of how the ordinary fermions
acquire mass. Many  extensions of technicolor have been suggested in
the literature to address this issue. Some of the extensions make
use of yet another strongly coupled gauge dynamics,  others
introduce fundamental scalars. It is even possible to marry
supersymmetry and technicolor.  Many variants of the schemes
presented above exist. A nice review of the major models is the one
by Hill and Simmons \cite{Hill:2002ap}. It is fair to say that at
the moment there is not yet a consensus on which is the correct ETC.
Although it is beyond the scope of this initial investigation to
provide a complete working scheme for mass generation we find it
instructive to construct the simplest model able to provide mass to
all of the fermions and which does not affect our results, but
rather improves them.

We parametrize our ETC, or better our ignorance about a complete ETC theory, with the (re)introduction of a single Higgs type doublet on the top of the minimal walking theory whose main purpose is to give mass to the ordinary fermions. This simple construction leads to no flavor changing neutral currents and does not upset the agreement with the precision tests which our MWT theory already passes brilliantly. We are able to give mass to all of the fermions and the contribution to the beta functions reads:
\begin{eqnarray}
b_3 & = & \frac{4}{3}N_g -11 \ ,\\
b_2 & = & \frac{4}{3} \left(
N_g + 1 \right) - \frac{22}{3}  + \frac{1}{6}\ ,\\
b_1 & = &  \frac{4}{3}\left( N_g+1 \right) +\frac{1}{10} \ ,
\end{eqnarray}
leading to
\begin{equation}
B_{theory}=0.71 \ ,
\end{equation}
a value which, at the one loop level, is even closer to the
experimental value of $0.72$ than the original MWT theory alone. The
unification scale is also slightly higher than in MWT alone and it
is of the order of $1.2 \times 10^{13}$~ GeV.  The ETC construction
presented above has already been used many times in the literature
\cite{Simmons:1988fu,Dine:1990jd,Kagan:1990az,Kagan:1991gh,Carone:1992rh,Carone:1993xc}.
We find the results very encouraging.  We wish to add that the need
for walking dynamics in the gauge sector is important since it helps
reducing the value of the S-parameter which is typically large even
before taking into account the problems due to the introduction of
an ETC sector.

\section{A New Extension of the Standard Model}
We wish to improve on the unification point (before taking into account of possible ETC type corrections) and delay it, energy-wise, to
avoid the experimental bounds on the proton decay.

We hence need a minimal modification of our extension of the SM with
the following properties: i) it is natural, i.e. it does not
reintroduce the hierarchy problem, ii) it does not affect the working
technicolor sector, iii) it allows for a straightforward unification
with a resulting theory which is asymptotically free, iv) it yields a
phenomenologically viable proton decay rate and possibly leads also
to dark matter candidates.

Point i) forces us to add new fermionic-type matter while ii) can be
satisfied by modifying the matter content of the SM per se. A simple
thing to do is to explore the case in which we consider adjoint
fermionic matter for the strong and weak interactions. We will show
that this is sufficient to greatly improve the proton decay problem
while also improving unification with respect to the MWT theory. To be
more specific, we add one colored
Weyl fermion transforming solely
according to the adjoint representation of $SU(3)$ and a Weyl fermion
transforming according to the adjoint
representation of $SU_L(2)$. These
fermions can
be identified with the gluino and wino in supersymmetric extensions of
the SM.  The big
hierarchy is still under control in the present model.

Since our theory is not supersymmetric the introduced fermions need
not be degenerate with the associated gauge bosons. Their masses can
be of the order of, or larger than, the electroweak scale. Finally,
naturality does not forbid the presence of a fermion associated to
the hypercharge gauge boson and hence this degree of freedom may
occur in the theory. Imagining a unification of the value of the
masses at the unification scale also requires the presence of such a
${U(1)}$ bino-type fermion.

In this case
the one-loop beta function coefficients are
\begin{eqnarray}
b_3 &=& \frac{4}{3}N_g -11 + 2 \ , \\
b_2 &=& \frac{4}{3} \big( N_g +1 \big) - \frac{22}{3} + \frac{4}{3} \ ,
\\
b_1 &=& \frac{4}{3} \big( N_g +1 \big) \ .
\end{eqnarray}
This gives $B_{\rm theory} = 13/18 \sim
0.72{(2)}$ which is in
excellent agreement with the experimental value. Note also that the
unification scale is ${M_{GUT} \sim 2.65 \times 10^{15} \text{GeV}}$
which brings the proton decay within the correct order of magnitude set by
experiments.
\begin{figure}[ht]
\begin{center}
\includegraphics[width=0.6\linewidth]{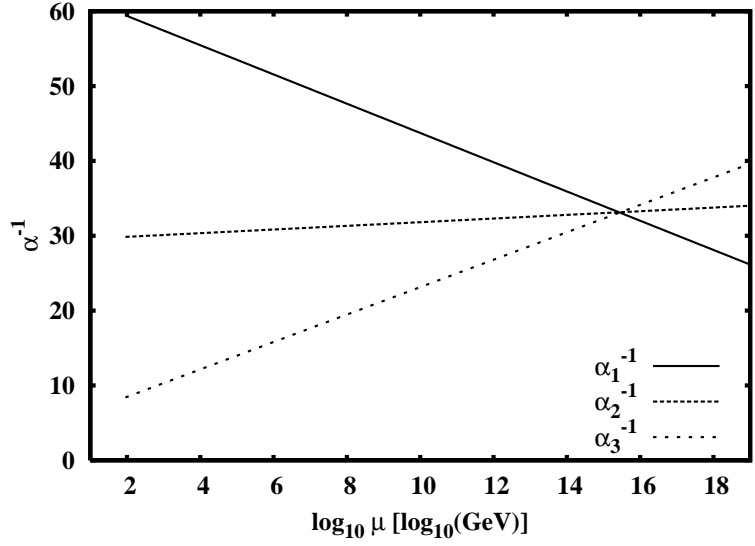}
\end{center}
\caption{The running of the three SM gauge couplings in the new model
  with also adjoint fermionic matter for the SM gauge groups.}
\label{SM-Improved}
\end{figure}

\begin{figure}[ht]
\begin{center}
\mbox{\subfigure{\resizebox{!}{5.3cm}{\includegraphics{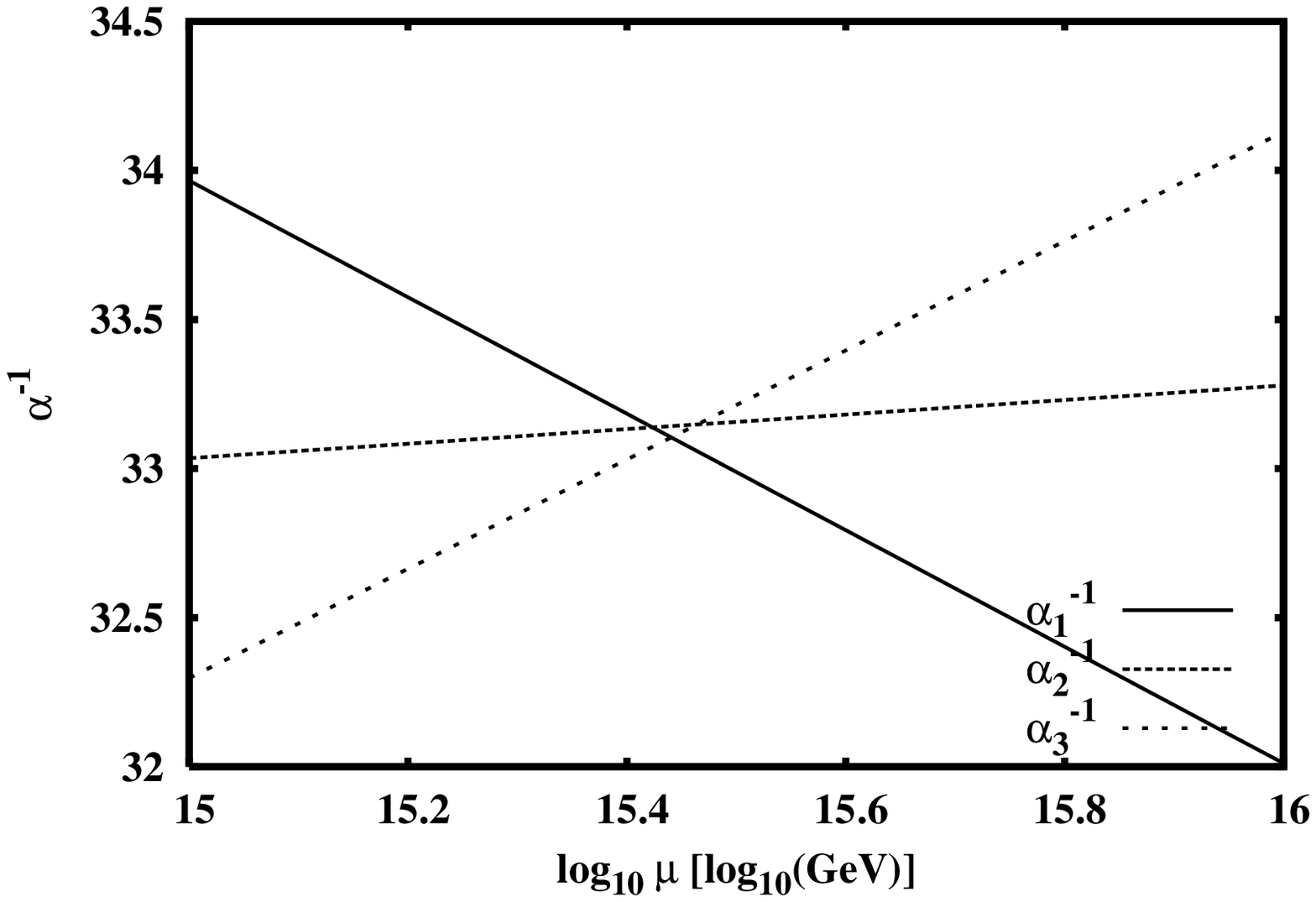}}}
\quad
\subfigure{\resizebox{!}{5.3cm}{\includegraphics{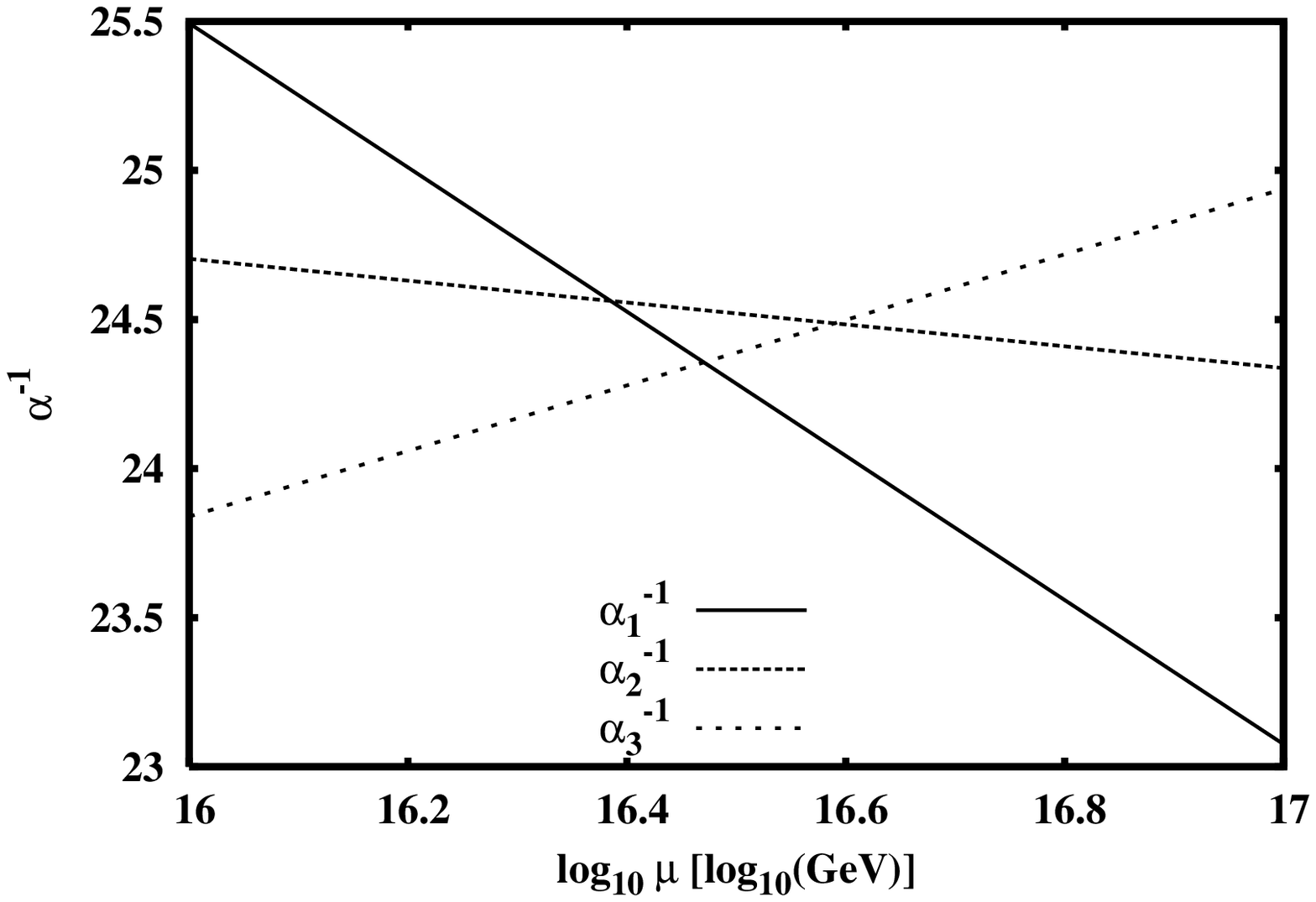}}}}
\end{center}
\caption{\textit{Left Panel}: A zoom around the unification point of the running of the
  three SM gauge couplings in the new model with extra fermionic adjoint matter
  for the SM gauge groups. \textit{Right Panel}: A zoom around the
  unification point for the couplings in the MSSM. }
\label{ZoomCompare}
\end{figure}



\subsection{Unifying Technicolor as Well}
We can make the technicolor coupling unify with the SM couplings, as
done in the MWT section.  Using
Eq.~(\ref{XX}), we find now $X\sim
10^8$ GeV. We recall here that $X$ is the scale above which our
technicolor theory becomes an ${SU(3)}$ gauge theory with the fermions
transforming according to the fundamental representation.

It is phenomenologically appealing that the scale $X$ is much higher than the electroweak scale. This allows our technicolor coupling to walk for a sufficiently large range of energy to allow for the introduction of extended technicolor interactions needed to give masses to the SM particles.
\begin{figure}[htbp]
\begin{center}
\includegraphics[width=0.6\linewidth]{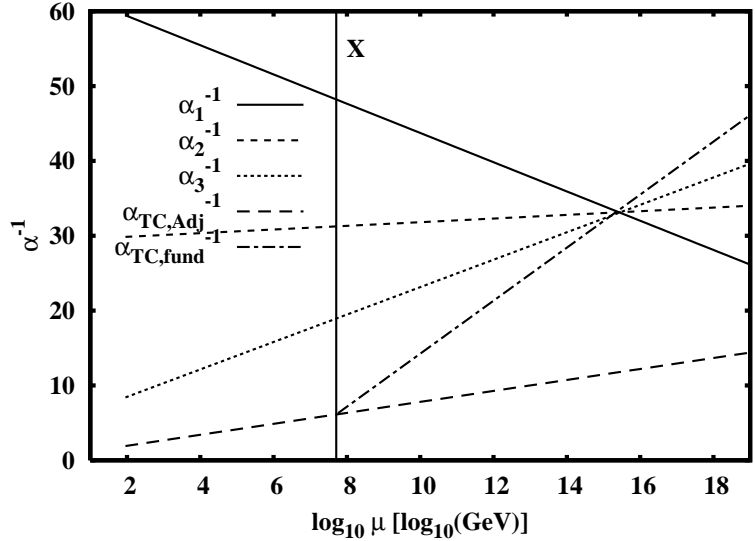}
\end{center}
\caption{The running of the three SM gauge couplings as well as the
  technicolor one. MWT is made to unify with the other three couplings
  by enhancing the gauge group from ${SU(2)}$ of technicolor
  to ${SU(3)}$ while
  keeping the same fermionic matter content. We see that the scale
  where this enhancement of the gauge group should dynamically occur
  to obtain complete unification is around $10^8$ GeV. }
\label{SM-ImprovedX}
\end{figure}
\subsection{Comparing with the MSSM and Hint of Dark Matter}
Unification of the SM gauge couplings is considered one of the
strongest points in favor of a supersymmetric extension of the SM
and hence it is reasonable to compare our results with the {SUSY}
ones. In {SUSY}, one finds $B_{\rm theory}=0.714$ which is
remarkably close to the experimental value $B_{\rm exp}\sim 0.72$
but it is not better than the value predicted in the present model
which is $0.72(2)$. Obviously this comparison must be taken with a
grain of salt since we still need to provide mass to the SM fermions
and take care of the threshold corrections. {}For example according
to the model introduced in section III, subsection G, to give masses
to all of the fermions yields a theoretical value for the
unification which is around $0.76$.

There are three possible candidates for dark matter here, depending on
which one is the lightest one and on the extended technicolor
interactions which we have not yet specified but that we will
explore in the future: The chargeless fermion in the adjoint
representation of $SU(2)_L$, i.e. the wino-like object as well as the
bino-type one.  The third possibility is the heavy neutrino-like fermion whose dark matter potential features are being currently investigated \cite{Kimmo}.

\section{Conclusions}
We have introduced a technicolor model which leads to the unification of the SM gauge couplings. At the one-loop level the model provides a higher degree of unification when compared to other technicolor models and to the minimal supersymmetric extension of the SM.

The phenomenology, both for collider experiments and cosmology, of the present extension of the SM is very
rich and needs to be explored in much detail.

The model has many
features in common with split and non-split supersymmetry \cite{Arkani-Hamed:2004fb,Giudice:2004tc} and also with very recent models proposed in \cite{Bajc:2006ia,Dorsner:2006fx} while others in common with
technicolor.

\acknowledgments
We thank T. Appelquist, B. Bajc, D.D. Dietrich, R. Foadi, M.T. Frandsen, C. Kouvaris and K. Tuominen for discussions or careful reading of the manuscript.

 The work of F.S. is supported by the Marie Curie Excellence
Grant under contract MEXT-CT-2004-013510.
F.S. is also supported as Skou fellow of the Danish Research
Agency.


\begin{thebibliography}{199}
\bibitem{TC}
S.~Weinberg,
``Implications Of Dynamical Symmetry Breaking: An Addendum,''
Phys.\ Rev.\ D {\bf 19}, 1277 (1979);
L.~Susskind,
``Dynamics Of Spontaneous Symmetry Breaking In The Weinberg-Salam Theory,''
Phys.\ Rev.\ D {\bf 20}, 2619 (1979).
\bibitem{Sannino:2004qp}
F.~Sannino and K.~Tuominen,
``Orientifold Theory Dynamics and Symmetry Breaking,'' Phys.\ Rev.\ D {\bf 71}, 051901 (2005).
arXiv:hep-ph/0405209.

\bibitem{Hong:2004td}
  D.~K.~Hong, S.~D.~H.~Hsu and F.~Sannino,
  ``Composite Higgs from higher representations,''
  Phys.\ Lett.\ B {\bf 597}, 89 (2004)
  [arXiv:hep-ph/0406200].

\bibitem{Dietrich:2005wk}
  D.~D.~Dietrich, F.~Sannino and K.~Tuominen,
  ``Light composite Higgs and precision electroweak measurements on the Z
  resonance: An update,''
  Phys.\ Rev.\ D {\bf 73}, 037701 (2006)
  [arXiv:hep-ph/0510217].

\bibitem{Dietrich:2005jn}
  D.~D.~Dietrich, F.~Sannino and K.~Tuominen,
  ``Light composite Higgs from higher representations versus electroweak
  precision measurements: Predictions for LHC,''
  Phys.\ Rev.\ D {\bf 72}, 055001 (2005)
  [arXiv:hep-ph/0505059].

\bibitem{Evans:2005pu}
  N.~Evans and F.~Sannino,
  ``Minimal walking technicolour, the top mass and precision electroweak
  measurements,''
  arXiv:hep-ph/0512080.



\bibitem{Gudnason:2006yj}
  S.~B.~Gudnason, C.~Kouvaris and F.~Sannino,
  ``Dark matter from new technicolor theories,''
  Phys.\ Rev.\ D {\bf 74} (2006) 095008
  [arXiv:hep-ph/0608055].
  S.~B.~Gudnason, C.~Kouvaris and F.~Sannino,
  ``Towards working technicolor: Effective theories and dark matter,''
  Phys.\ Rev.\ D {\bf 73}, 115003 (2006)
  [arXiv:hep-ph/0603014].

\bibitem{Holdom:1981rm}
B.~Holdom, ``Raising The Sideways Scale,''
Phys.\ Rev.\ D {\bf 24}, 1441 (1981).

\bibitem{Yamawaki:1985zg}
K.~Yamawaki, M.~Bando and K.~i.~Matumoto,
``Scale Invariant Technicolor Model And A Technidilaton,''
Phys.\
Rev.\ Lett.\ {\bf 56}, 1335 (1986).

\bibitem{Appelquist:an}
T.~W.~Appelquist, D.~Karabali and L.~C.~R.~Wijewardhana,
``Chiral Hierarchies And The Flavor Changing Neutral
Current Problem In Technicolor,''
Phys.\ Rev.\ Lett.\ {\bf 57}, 957 (1986).

\bibitem{MY}
  V.~A.~Miransky and K.~Yamawaki,
  ``Conformal phase transition in gauge theories,''
  Phys.\ Rev.\ D {\bf 55}, 5051 (1997)
  [Erratum-ibid.\ D {\bf 56}, 3768 (1997)]
  [arXiv:hep-th/9611142].

  V.~A.~Miransky, T.~Nonoyama and K.~Yamawaki,
  ``On The Phase Diagram Of Asymptotically Free Gauge Theories With Additional
  Four Fermion Interaction,''
  Mod.\ Phys.\ Lett.\ A {\bf 4}, 1409 (1989).

\bibitem{Lane:1989ej}
K.~D.~Lane and E.~Eichten,
``Two Scale Technicolor,''
Phys.\ Lett.\ B {\bf 222}, 274 (1989).
E.~Eichten and K.~D.~Lane,
``Dynamical Breaking Of Weak Interaction Symmetries,''
Phys.\ Lett.\ B {\bf 90}, 125
(1980).


\bibitem{Cohen:1988sq}
  A.~G.~Cohen and H.~Georgi,
  ``Walking Beyond The Rainbow,''
  Nucl.\ Phys.\ B {\bf 314}, 7 (1989).


\bibitem{Peskin:1990zt}
  M.~E.~Peskin and T.~Takeuchi,
  ``A New Constraint On A Strongly Interacting Higgs Sector,''
  Phys.\ Rev.\ Lett.\  {\bf 65}, 964 (1990).


\bibitem{Appelquist:1998xf}
T.~Appelquist and F.~Sannino,
``The physical spectrum of conformal SU(N) gauge theories,''
Phys.\ Rev.\ D {\bf 59}, 067702 (1999) [arXiv:hep-ph/9806409].

\bibitem{Sundrum:1991rf}
  R.~Sundrum and S.~D.~H.~Hsu,
  ``Walking technicolor and electroweak radiative corrections,''
  Nucl.\ Phys.\ B {\bf 391}, 127 (1993)
  [arXiv:hep-ph/9206225].

\bibitem{Appelquist:1999dq}
T.~Appelquist, P.~S.~Rodrigues da Silva and F.~Sannino,
``Enhanced global symmetries and the chiral phase transition,''
Phys.\ Rev.\ D {\bf 60}, 116007 (1999) [arXiv:hep-ph/9906555].
Z.~y.~Duan, P.~S.~Rodrigues da Silva and F.~Sannino,
``Enhanced global symmetry constraints on epsilon terms,''
Nucl.\ Phys.\ B {\bf 592}, 371 (2001) [arXiv:hep-ph/0001303].

\bibitem{Dietrich:2006cm}
  D.~D.~Dietrich and F.~Sannino,
 "Conformal window of SU(N) gauge theories with fermions in higher dimensional representations".
  Phys.\ Rev.\  D {\bf 75}, 085018 (2007)
  [arXiv:hep-ph/0611341].


\bibitem{Witten:fp}
E.~Witten,
``An SU(2) Anomaly,''
Phys.\ Lett.\ B {\bf 117}, 324 (1982).

\bibitem{Simmons:1988fu}
  E.~H.~Simmons,
  "Phenomenology of a technicolor model with heavy scalar doublet"
  Nucl.\ Phys.\  B {\bf 312}, 253 (1989).

\bibitem{Dine:1990jd}
  M.~Dine, A.~Kagan and S.~Samuel,
  ``Naturalness in supersymmetry, or raising the supersymmetry breaking scale,''
  Phys.\ Lett.\  B {\bf 243}, 250 (1990).

\bibitem{Kagan:1990az}
  A.~Kagan and S.~Samuel,
  ``The Family mass hierarchy problem in bosonic technicolor,''
  Phys.\ Lett.\  B {\bf 252}, 605 (1990).

\bibitem{Kagan:1991gh}
  A.~Kagan and S.~Samuel,
  ``Renormalization group aspects of bosonic technicolor,''
  Phys.\ Lett.\  B {\bf 270}, 37 (1991).

\bibitem{Carone:1992rh}
  C.~D.~Carone and E.~H.~Simmons,
  ``Oblique corrections in technicolor with a scalar,''
  Nucl.\ Phys.\  B {\bf 397}, 591 (1993)
  [arXiv:hep-ph/9207273].

\bibitem{Carone:1993xc}
  C.~D.~Carone and H.~Georgi,
  ``Technicolor with a massless scalar doublet,''
  Phys.\ Rev.\  D {\bf 49}, 1427 (1994)
  [arXiv:hep-ph/9308205].




\bibitem{Yao:2006px}
  W.~M.~Yao {\it et al.}  [Particle Data Group],
  ``Review of particle physics,''
  J.\ Phys.\ G {\bf 33}, 1 (2006).

\bibitem{Li:2003zh}
  L.~F.~Li and F.~Wu,
  ``Coupling constant unification in extensions of standard model,''
  Int.\ J.\ Mod.\ Phys.\ A {\bf 19}, 3217 (2004)
  [arXiv:hep-ph/0304238].


\bibitem{Christensen:2005bt}
  N.~D.~Christensen and R.~Shrock,
  ``On the unification of gauge symmetries in theories with dynamical  symmetry breaking,''
  Phys.\ Rev.\ D {\bf 72}, 035013 (2005)
  [arXiv:hep-ph/0506155].


\bibitem{Christensen:2005cb}
  N.~D.~Christensen and R.~Shrock,
  ``Technifermion representations and precision electroweak constraints,''
  Phys.\ Lett.\ B {\bf 632}, 92 (2006)
  [arXiv:hep-ph/0509109].

\bibitem{Hill:2002ap}
  C.~T.~Hill and E.~H.~Simmons,
  Phys.\ Rept.\  {\bf 381}, 235 (2003)
  [Erratum-ibid.\  {\bf 390}, 553 (2004)]
  [arXiv:hep-ph/0203079].

\bibitem{Giudice:2004tc}
  G.~F.~Giudice and A.~Romanino,
  ``Split supersymmetry,''
  Nucl.\ Phys.\ B {\bf 699}, 65 (2004)
  [Erratum-ibid.\ B {\bf 706}, 65 (2005)]
  [arXiv:hep-ph/0406088].


\bibitem{Aoki:1999tw}
  S.~Aoki {\it et al.}  [JLQCD Collaboration],
  ``Nucleon decay matrix elements from lattice QCD,''
  Phys.\ Rev.\ D {\bf 62} (2000) 014506
  [arXiv:hep-lat/9911026].

\bibitem{Ross:1985ai}
  G.~G.~Ross,
  ``Grand Unified Theories,'' Frontiers in Physics, ABP.


\bibitem{Suzuki:2001rb}
  Y.~Suzuki {\it et al.}  [TITAND Working Group],
  ``Multi-Megaton water Cherenkov detector for a proton decay search:  TITAND
  arXiv:hep-ex/0110005.



\bibitem{Arkani-Hamed:2004fb}
  N.~Arkani-Hamed and S.~Dimopoulos,
  ``Supersymmetric unification without low energy supersymmetry and  signatures for fine-tuning at the LHC,''
  JHEP {\bf 0506}, 073 (2005)
  [arXiv:hep-th/0405159].

\bibitem{Bajc:2006ia}
  B.~Bajc and G.~Senjanovic,
  ``Seesaw at LHC,''
  arXiv:hep-ph/0612029.
\bibitem{Dorsner:2006fx}
  I.~Dorsner and P.~F.~Perez,
  ``On fermion masses, unification, proton decay and the UV cutoff of adjoint
  SU(5),''
  arXiv:hep-ph/0612216.

\bibitem{Kimmo}
Private communication with Kimmo Tuominen.











\end{thebibliography}
\end{document}